\documentclass[journal]{IEEEtran}
\usepackage{graphicx}
\usepackage{amsmath}
\usepackage{amssymb}
\usepackage{cite}
\usepackage{psfig}
\usepackage{subfigure}
\usepackage{booktabs}
\usepackage{pseudocode}
\newenvironment{squishitemize}
  {\begin{list}{\textasteriskcentered}{%
    \setlength{\itemsep}{0pt}%
    \setlength{\parsep}{0pt}%
    \setlength{\topsep}{0pt}%
    \setlength{\parskip}{0pt} %
    \setlength{\labelwidth}{2.5in}%
    \setlength{\labelsep}{0.1in} %
    \setlength{\leftmargin}{0.37in} %
    }}
  {\end{list}}

\newcounter{mytempeqncnt}

\begin{document}

\title{Symbol and Bit Mapping Optimization for Physical-Layer Network Coding with Pulse Amplitude Modulation}

\author{Ronald Y. Chang,~\IEEEmembership{Member,~IEEE}, Sian-Jheng Lin, and Wei-Ho Chung,~\IEEEmembership{Member,~IEEE}%
\thanks{Manuscript received October 3, 2012; revised March 3, 2013 and May 23, 2013; accepted June 14, 2013. The associate editor coordinating the review of this paper and approving it for publication was M. Ardakani.}%
\thanks{This work was supported in part by the National Science Council, Taiwan, under grants NSC 102-2218-E-001-001, NSC 101-2221-E-001-002, NSC 101-2221-E-001-008, NSC 101-2622-E-001-001-CC3, and 32T-1010721-1C.}%
\thanks{R. Y. Chang (corresponding author), S.-J. Lin, and W.-H. Chung are with the Research Center for Information Technology Innovation, Academia Sinica, Taipei, Taiwan (email: \{rchang, sjlin, whc\}@citi.sinica.edu.tw).}}

\markboth{IEEE Transactions on Wireless Communications, Accepted for Publication}%
{Chang \MakeLowercase{\textit{et al.}}: Symbol and Bit Mapping Optimization for Physical-Layer Network Coding with Pulse Amplitude Modulation}

\maketitle


\begin{abstract}
In this paper, we consider a two-way relay network in which two users exchange messages through a single relay using a physical-layer network coding (PNC) based protocol. The protocol comprises two phases of communication. In the multiple access (MA) phase, two users transmit their modulated signals concurrently to the relay, and in the broadcast (BC) phase, the relay broadcasts a network-coded (denoised) signal to both users. Nonbinary and binary network codes are considered for uniform and nonuniform pulse amplitude modulation (PAM) adopted in the MA phase, respectively. We examine the effect of different choices of symbol mapping (i.e., mapping from the denoised signal to the modulation symbols at the relay) and bit mapping (i.e., mapping from the modulation symbols to the source bits at the user) on the system error-rate performance. A general optimization framework is proposed to determine the optimal symbol/bit mappings with joint consideration of noisy transmissions in both communication phases. Complexity-reduction techniques are developed for solving the optimization problems. It is shown that the optimal symbol/bit mappings depend on the signal-to-noise ratio (SNR) of the channel and the modulation scheme. A general strategy for choosing good symbol/bit mappings is also presented based on a high-SNR analysis, which suggests using a symbol mapping that aligns the error patterns in both communication phases and Gray and binary bit mappings for uniform and nonuniform PAM, respectively.
\end{abstract}

\IEEEpeerreviewmaketitle


\section{Introduction} \label{sec:intro}

\IEEEPARstart{I}{n} a two-way relay network, the denoise-and-forward (DNF) protocol with physical-layer network coding (PNC) \cite{ZhangLiew06,PopovskiYomo06,PopovskiYomo07} uses two time slots to complete information exchange between two users through a single relay. The two users transmit concurrently to the relay in the first time slot (the multiple access (MA) phase) and the relay performs denoising on the received interfered signal and broadcasts a denoised signal in the second time slot (the broadcast (BC) phase) to enable each user to decode each other's information at its side. Denoising is a many-to-one mapping technique that maps the complex field operation in the wireless channel to the finite field operation for network coding. The DNF protocol increases the achievable throughput at high signal-to-noise ratio (SNR) \cite{PopovskiYomo06} compared to the conventional decode-and-forward (DF) relaying (i.e., a four-phase protocol) and DF relaying with network coding (NC) (i.e., a three-phase protocol \cite{WuChou04}) due to the improved time efficiency. The DNF protocol achieves higher throughput at low SNR \cite{PopovskiYomo06} than the amplify-and-forward (AF) relaying or analog network coding \cite{KattiGollakota07,SongLi10} due to the avoidance of noise amplification. The capacity-achieving lattice coding scheme \cite{ErezZamir04,ErezLitsyn05,NazerGastpar11_2,WilsonNarayanan10,NamChung08} is developed from an information-theoretic perspective where users' messages are encoded into high-dimensional lattice codes and transmitted to the relay. Extensions and generalizations have been developed for systems with multiantenna nodes \cite{DingKrikidis11} and asynchronous scenarios \cite{ZhangLiew06_2,LuLiew12,RossettoZorzi09}. For a more extensive coverage of the two-way relay network with PNC, see the tutorial by Nazer and Gastpar \cite{NazerGastpar11} as well as Liew {\it et al.} \cite{LiewZhang11}.

The denoising scheme for the two-way relay network depends on the modulation adopted for transmission in the MA phase. When the two users transmit binary phase-shift keying (BPSK) signals, a simple binary denoising based on exclusive-or (XOR) operation can be performed. The XOR-based denoising can also be used for quadrature phase-shift keying (QPSK), since QPSK is a pair of orthogonal BPSK. In the case of QPSK and higher-order modulations, however, the denoise mapper depends on the channel gains between the users and the relay. The optimal denoise mapping for QPSK under asymmetric channel gains was derived by Koike-Akino {\it et al.} \cite{KoikeAkinoPopovski09} based on the design strategy of finding a denoise mapping such that the minimum Euclidean distance between the received signals at the relay associated with distinct network codes (denoised signals) is maximized. The denoise mapper for general squared quadrature amplitude modulation (QAM) signals was studied by Namboodiri and Rajan \cite{NamboodiriRajan12} for Gaussian channels with asymmetric gains. Binary coded denoising for general PSK was investigated by Noori and Ardakani \cite{NooriArdakani12}, where legitimate bit mappings for PSK symbols that avoid mapping ambiguity at the relay and thus enable binary denoising are determined for Gaussian channels with symmetric gains. The denoise mapper for general PSK was also studied by Muralidharan {\it et al.} \cite{MuralidharanNamboodiri12}. Faraji-Dana and Mitran \cite{FarajiDanaMitran13} studied unconventional $q$-PSK modulation ($q=2,3,4,5$) for channel-coded two-way relaying with PNC, and investigated the error-rate performance of different mappings from GF($q$) to the $q$-PSK constellation. Denoising for pulse amplitude modulation (PAM) was investigated by Yang {\it et al.} \cite{YangChoi10}, where the spacings between higher-order PAM constellation points are modified (i.e., nonuniform PAM) to enable the use of binary denoising without mapping ambiguity at the relay, and the optimal bit mapping of PAM symbols is determined numerically through an exhaustive computer search. Network coded modulation where NC and modulation are jointly designed was proposed for PSK, PAM, and QAM for two-way relaying \cite{SykoraBurr10,ChenHanzo11}.

In this paper, we propose an analytical formulation of symbol and bit mapping optimization for the DNF protocol given some predetermined nonbinary or binary denoise mapper for uniform (conventional) or nonuniform higher-order PAM adopted in the MA phase. We consider symmetric channel gains and perfect synchronization between signals transmitted by the two users. For each denoise mapper, the design freedom in choosing the symbol mapping (i.e., mapping from the denoised signal to the modulation symbols at the relay) and the bit mapping (i.e., mapping from the modulation symbols to the source bits at the user) motivates the search for the optimal symbol and/or bit mappings that yield the optimal system error-rate performance. The main contributions and findings of this paper are summarized below:
\begin{itemize}
\item We propose a general symbol and bit mapping optimization framework that jointly considers the noisy transmission in both MA and BC phases. Complexity-reduction methods are suggested to ease the complexity of solving the optimization problems.
\item We observe that optimal mappings depend on the channel SNR and the modulation. We present optimal mapping results for uniform/nonuniform 4/8-PAM adopted in the MA phase. A high-SNR analysis is conducted to interpret the results and offer a good rule of thumb for the mapping design: select a symbol mapping that aligns the most likely error patterns in MA and BC phases, and use Gray and binary bit mappings for uniform and nonuniform PAM, respectively.
\end{itemize}

The outline of this paper is as follows. Sec.~\ref{sec:system} presents the system description. The proposed general optimization framework is described in Sec.~\ref{sec:method}. Performance results and discussions are presented in Sec.~\ref{sec:simulation}. Conclusion is given in Sec.~\ref{sec:conclusion}.

\section{System Description} \label{sec:system}

\begin{figure}[tb]
\begin{center}
\includegraphics[width=0.5\columnwidth]{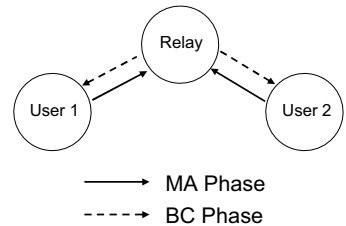}
\caption{A wireless two-way relay network with PNC.} \label{fig:network}
\end{center}
\end{figure}

We consider a two-way relay network with two users wishing to achieve information exchange and one relay. Direct communication is assumed infeasible and thus users can only communicate through the relay. The communication adopts the PNC technique and takes place in two phases, i.e., multiple access (MA) and broadcast (BC) phases, as shown in Fig.~\ref{fig:network}. Higher-order PAM (i.e., 4-PAM or 8-PAM) is considered for all transmissions. The two users and the relay each have a single antenna. For simplicity, we assume unit channel gains for all links and perfect synchronization between signals transmitted by the two users in the MA phase.

{\it The MA phase:} In the MA phase, each user $i$ ($i=1,2$) transmits the modulated signal $X_i=M_Q(S_i)$ simultaneously to the relay through additive white Gaussian noise (AWGN) channels, where $M_Q:{\mathbb Z}_Q\mapsto {\cal A}_Q^{eq} \mbox{ or } {\cal A}_Q^{neq}$ is the constellation mapper at the users which adopts a natural-order mapping from the source symbol $S_i\in {\mathbb Z}_Q=\{0,1,2,\ldots,Q-1\}$ to the uniform or nonuniform $Q$-PAM alphabet (denoted by ${\cal A}_Q^{eq}$ or ${\cal A}_Q^{neq}$). The source symbol $S_i$ is selected equiprobably from ${\mathbb Z}_Q$ and each distinct symbol is associated with a unique binary sequence of length $q=\log_2 Q$, denoted by $B_Q(S_i)=B_{i1} \, B_{i2} \,\cdots\, B_{iq}$, where $B_Q$ is the bit mapper. The received signal at the relay is given by
\begin{equation}
Y_R = X_1 + X_2 + Z_R
\end{equation}
where $Z_R$ is Gaussian noise with variance $\sigma^2$. 

After receiving the noisy combined signal from the two users, the relay performs maximum-likelihood (ML) detection on $Y_R$ to obtain the estimated $S_1$ and $S_2$, i.e.,
\begin{equation} \label{eq:ML}
\big(\hat{S}_1,\hat{S}_2\big)=\underset{(s_1,s_2)\in {\mathbb Z}_Q\times {\mathbb Z}_Q}{\operatorname{argmin}} \left|Y_R-\Big(M_Q(s_1)+M_Q(s_2)\Big)\right|^2.
\end{equation}
Note that the optimal ordered pair $(\hat{S}_1,\hat{S}_2)$ in (\ref{eq:ML}) is not unique; specifically, $(\hat{S}_1,\hat{S}_2)$ and $(\hat{S}_2,\hat{S}_1)$ will produce the same likelihood metric. Thus, joint decoding of the two users' messages at the relay is infeasible. The relay instead adopts a many-to-one mapping technique called denoising to enable successful decoding at the user side. The denoise mapper (or network code) $C:{\mathbb Z}_Q^2\mapsto {\mathbb Z}_Q$ must meet the well-known ``exclusive law" to ensure unique decodability at the user side \cite{KoikeAkinoPopovski09,PopovskiYomo06}:
\begin{squishitemize}
\item[1)] $C(S_1,S_2)\neq C(S_1',S_2)$ for any $S_1\neq S_1'\in {\mathbb Z}_Q$ and $S_2 \in {\mathbb Z}_Q$;
\item[2)] $C(S_1,S_2)\neq C(S_1,S_2')$ for any $S_2\neq S_2'\in {\mathbb Z}_Q$ and $S_1 \in {\mathbb Z}_Q$.
\end{squishitemize}
Based on $\hat{S}_1$ and $\hat{S}_2$, the relay obtains the denoised signal $\hat{U}=C(\hat{S}_1,\hat{S}_2)$. 

{\it The BC phase:} In the BC phase, the relay broadcasts the modulated denoised signal $X_R=M_{R,Q}(\hat{U})$ to both users, where $M_{R,Q}:{\mathbb Z}_Q\mapsto {\cal A}_Q^{eq}$ is the constellation mapper at the relay. Only uniform PAM is considered at the relay, which may or may not be the same as the modulation employed at the users. The received signal at user $i$ ($i=1,2$) is given by
\begin{equation}
Y_i = X_R + Z_i
\end{equation}
where $Z_i$ is Gaussian noise with variance $\sigma^2$. 

For decoding the other user's information, each user $i$ performs ML detection on $Y_i$ to obtain the estimated $X_R$ and (after demodulation) $\hat{U}$. User 2's signal detected at user 1 is given by
\begin{equation} \label{eq:ML2}
\tilde{S}_2=\underset{s\in {\mathbb Z}_Q}{\operatorname{argmin}} \left|Y_1-M_{R,Q}\big(C(S_1, s)\big)\right|^2.
\end{equation}
User 1's signal detected at user 2 can be formulated similarly. Due to the established properties of $C$, each user can uniquely decode each other user with its own information.

In the following, we briefly describe how uniform or nonuniform PAM adopted in the MA phase results in different superposed constellations and consequently different denoise mapping strategies at the relay.

\begin{figure}[tb]
\begin{center}
\includegraphics[width=0.45\columnwidth]{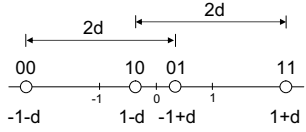}
\caption{General 4-PAM constellations at the users.} \label{fig:4PAM_users}
\end{center}
\end{figure}

\begin{figure}[tb]
\begin{center}
\includegraphics[width=\columnwidth]{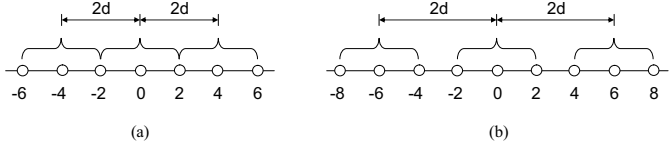}
\caption{Superposed constellation points at the relay for (a) uniform 4-PAM ($d=2$) and (b) nonuniform 4-PAM ($d=3$) adopted in the MA phase.} \label{fig:4PAM}
\end{center}
\end{figure}

{\it Uniform 4-PAM:} To facilitate our discussion on uniform and nonuniform 4-PAM, consider without loss of generality a specific bit mapping $B_4(S_i)=B_{i1} \, B_{i2}$ so that the transmitted signal $X_i=M_4(S_i)$ for user $i$ ($i=1,2$) can be expressed as
\begin{equation} \label{eq:X_i}
X_i=M_2(B_{i1})+d\times M_2(B_{i2})
\end{equation}
where $M_2(B)=1$ if $B=1$, and $-1$ otherwise, and $2d$ ($d>1$) represents the distance between constellation points labeled by the same bit $B_{i1}$, as shown in Fig.~\ref{fig:4PAM_users}. The combined noiseless signal at the relay is given by
\begin{equation} \label{eq:Y_R}
Y_R=\underbrace{\Big(M_2(B_{11})+M_2(B_{21})\Big)}_{\in\{-2,0,2\}}+d\times\underbrace{\Big(M_2(B_{12})+M_2(B_{22})\Big)}_{\in\{-2,0,2\}}.
\end{equation}
The expression in (\ref{eq:Y_R}) suggests that the superposed constellation points at the relay consist of three copies of $\{-2,0,2\}$ shifted by $-2d$, 0, and $2d$, respectively. The uniform 4-PAM adopts $d=2$ in (\ref{eq:X_i}), resulting in the constellation points ${\cal A}_4^{eq}=\{-3, -1, 1, 3\}$. The superposed constellation points at the relay are shown in Fig.~\ref{fig:4PAM}(a). A feasible denoise mapper for uniform $Q$-PAM corresponds to the modulo-$Q$ addition over the reals \cite{WilsonNarayanan10}, i.e.,
\begin{equation} \label{eq:denoise_mod}
C(S_1,S_2)=\big[S_1+S_2\big]\operatorname{mod} Q.
\end{equation}

{\it Nonuniform 4-PAM:} The denoise mapper in (\ref{eq:denoise_mod}) is a nonbinary PNC for $Q>2$. It is sometimes useful to consider a binary PNC-based denoise mapper \cite{NazerGastpar11}, i.e.,
\begin{equation} \label{eq:denoise_xor}
C(S_1,S_2)=S_1\oplus_q S_2
\end{equation}
where $\oplus_q$ represents bit-wise XOR operation\footnote{Note that the bit-wise XOR operation here is based on the ``natural" binary representation of $S_i$ (e.g., $000$ for $S_i=0$, $010$ for $S_i=2$, $100$ for $S_i=4$, etc., when $Q=8$), rather than $B_Q(S_i)$.}, or equivalently, the addition over GF($Q$) or GF($2^q$). The denoise mapper in (\ref{eq:denoise_xor}) however cannot be used for uniform $Q$-PAM due to the ambiguity problem at certain received signal levels at the relay \cite{YangChoi10}. In order to use the denoise mapper in (\ref{eq:denoise_xor}) it has been proposed \cite{YangChoi10} that the spacings between $Q$-PAM constellation points be modified. The proposed nonuniform 4-PAM \cite{YangChoi10}, in our description, adopts $d=3$ instead of $d=2$ in (\ref{eq:X_i}), resulting in the constellation points ${\cal A}_4^{neq}=\{-4,-2,2,4\}$. This modification separates the three copies of $\{-2,0,2\}$ at the relay, as shown in Fig.~\ref{fig:4PAM}(b). Since each superposed constellation point uniquely corresponds to one of the nine combinations of $M_2(B_{11})+M_2(B_{21})\in \{-2,0,2\}$ and $M_2(B_{12})+M_2(B_{22})\in \{-2,0,2\}$ in (\ref{eq:Y_R}), the denoise mapping can now be done at the bit level and the denoise mapper in (\ref{eq:denoise_xor}) can be used. Note that the nonuniform arrangement inevitably reduces the minimum constellation spacing (after power normalization) and leads to higher average bit-error-rate (BER) at the relay.

{\it Uniform and nonuniform 8-PAM:} For 8-PAM, the transmitted signal $X_i=M_8(S_i)$ for user $i$ ($i=1,2$) can be expressed as
\begin{equation}
X_i=M_2(B_{i1})+d\times M_2(B_{i2})+d^2\times M_2(B_{i3})
\end{equation}
where $d=2$ for uniform 8-PAM and $d=3$ for nonuniform 8-PAM. Effectively, the constellation points for uniform and nonuniform 8-PAM are ${\cal A}_8^{eq}=\{-7,-5,-3,-1,1,3,5,7\}$ and ${\cal A}_8^{neq}=\{-13,-11,-7,-5,5,7,11,13\}$, respectively. The superposed constellation at the relay has 15 levels for uniform 8-PAM and 27 levels for nonuniform 8-PAM. The denoise mappers in (\ref{eq:denoise_mod}) and (\ref{eq:denoise_xor}) are used for uniform and nonuniform 8-PAM, respectively, with $Q=8$.

\section{The Proposed Symbol and Bit Mapping Optimization Framework} \label{sec:method}

In the previous section we describe the DNF two-way relaying mechanism without specifying $M_{R,Q}$, the constellation/symbol mapper at the relay, and $B_Q$, the bit mapper at the users. (Note that the symbol mapper at the users, $M_Q$, adopts a natural-order mapping and is thus determined, and the bit mapper is not employed at the relay since the relay does not perform decoding itself.) In this section, we examine how different relay symbol mappings and user bit mappings might affect the decoding symbol-error-rate (SER) and BER performance at the users, and propose an optimization framework for finding the optimal symbol/bit mappings. It is worthwhile to mention that the proposed framework can be applied to the case of QAM since QAM signals can be viewed as two parallel PAM signals.

\begin{table*}[tb]
\begin{center}
\caption{Two Different Relay Symbol Mappings $M_{R,4}$ When Uniform 4-PAM Is Used in the MA Phase}
\label{tab:symbolmapping} \vspace*{1mm}
\begin{tabular}{|r||ccccccc|}
\hline
$M_4(S_1)+M_4(S_2)$ & $-6$ & $-4$ & $-2$ & $0$ & $2$ & $4$ & $6$ \\\hline
$(S_1,S_2)$ & $(0,0)$ & $(0,1)$ & $(0,2)$ & $(0,3)$ & $(1,3)$ & $(2,3)$ & $(3,3)$ \\
 & & $(1,0)$ & $(2,0)$ & $(3,0)$ & $(3,1)$ & $(3,2)$ & \\
 & & & $(1,1)$ & $(1,2)$ & $(2,2)$ & & \\
 & & & & $(2,1)$ & & & \\\hline
$U=C(S_1,S_2)$ & $0$ & $1$ & $2$ & $3$ & $0$ & $1$ & $2$ \\\hline
1) $M_{R,4}(U)$ & $-3$ & $-1$ & $1$ & $3$ & $-3$ & $-1$ & $1$ \\
2) $M_{R,4}(U)$ & $-3$ & $1$ & $-1$ & $3$ & $-3$ & $1$ & $-1$ \\\hline
\end{tabular}
\end{center}
\end{table*}

\begin{figure*}[tb]
\begin{center}
\includegraphics[width=1.3\columnwidth]{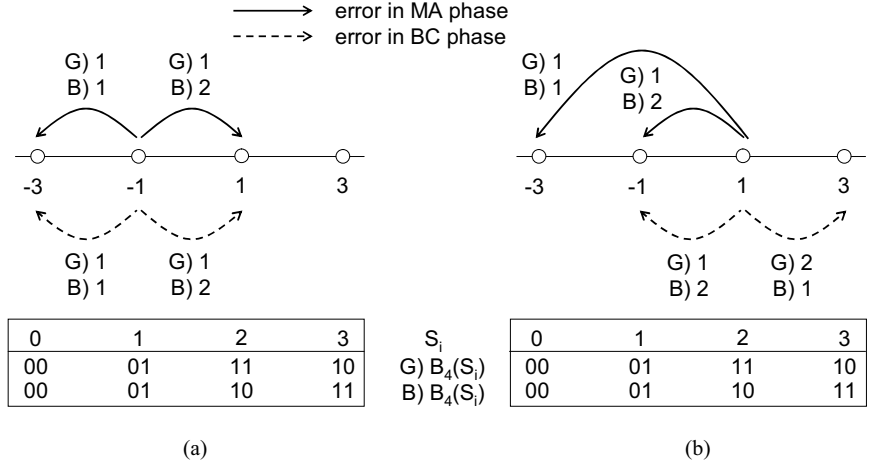}
\caption{Dominant error patterns when the transmitted source symbol pair is $(S_1,S_2)=(0,1)$, for (a) symbol mapping 1 and (b) symbol mapping 2 in Table \ref{tab:symbolmapping}. Each error pattern represents one symbol error and various numbers of bit errors (labeled next to the line) for Gray (denoted by ``G") and binary (denoted by ``B") bit mappings at the user.} \label{fig:errorpatterns}
\end{center}
\end{figure*}

\subsection{An Illustrative Example} \label{sec:method1}

In this example, we consider uniform 4-PAM being adopted in both MA and BC phases. Assume that the actual transmitted source symbol pair is $(S_1,S_2)=(0,1)$, i.e., user 1 transmits $-3$ and user 2 transmits $-1$ after modulation in the MA phase. The received signal at the relay is therefore $Y_R=-4+Z_R$. Due to noise perturbation, the relay might erroneously detect a superposed constellation point other than $-4$, which will most likely be the adjacent constellation point $-6$ or $-2$ when the noise is moderate. The effect of this noise-induced error in the MA phase on the transmit signal in the BC phase depends on the symbol mapper $M_{R,4}$. Specifically, if symbol mapping 1 in Table \ref{tab:symbolmapping} is used, the relay will broadcast $-3$ or $1$ when it should broadcast $-1$ if no error occurred. If symbol mapping 2 in Table \ref{tab:symbolmapping} is used, the relay will broadcast $-3$ or $-1$ when it should broadcast $1$ if no error occurred. These dominant error patterns in the MA phase are illustrated by solid lines in Fig.~\ref{fig:errorpatterns} for the two symbol mappings. As can be seen, different symbol mappings affect the distribution of error patterns. Especially, a dominant error that occurs among adjacent superposed constellation points at the relay will produce an effective error that may or may not occur among adjacent constellation points at the user side.

Due to noisy BC channels, each user might erroneously detect a constellation point other than the one transmitted from the relay. The dominant error patterns in the BC phase are illustrated by dashed lines in Fig.~\ref{fig:errorpatterns} for the two symbol mappings assuming there was no error in the MA phase. As can be seen, a dominant error will occur among adjacent constellation points at the user side.

A decoding symbol error at the user side may be due to errors in the MA phase or in the BC phase, or both. For example, if the dominant error pattern $-1\rightarrow -3$ or $-1\rightarrow 1$ in Fig.~\ref{fig:errorpatterns}(a) occurs, user 1 will erroneously decode user 2's signal as $\tilde{S}_2=0$ or $\tilde{S}_2=2$, respectively. If the dominant error pattern $1\rightarrow -3$, $1\rightarrow -1$, or $1\rightarrow 3$ in Fig.~\ref{fig:errorpatterns}(b) occurs, user 1 will erroneously decode user 2's signal as $\tilde{S}_2=0$, $\tilde{S}_2=2$, or $\tilde{S}_2=3$, respectively. The number of decoding bit errors depends on the bit mapper $B_4$. The number of bit errors associated with the symbol error from $S_2$ to $\tilde{S}_2$ is given by the Hamming distance between the bit sequences that represent $S_2$ and $\tilde{S}_2$, denoted by $d_H\big(B_4(S_2),B_4(\tilde{S}_2)\big)$. For example, if symbol mapping 1 is used, the number of bit errors for error pattern $-1\rightarrow -3$ is given by $d_H\big(B_4(1),B_4(0)\big)$, which is equal to 1 for both Gray and binary mappings, as shown in Fig.~\ref{fig:errorpatterns}(a). The number of bit errors for error pattern $-1\rightarrow 1$ is given by $d_H\big(B_4(1),B_4(2)\big)$, which is equal to 1 for Gray mapping and 2 for binary mapping.

As illustrated in this example, different symbol and bit mappings might affect the SER and BER decoding performance at the user side. It is however not immediately clear as to the sum effect of noisy transmissions in MA and BC phases on the combined error patterns and on the decoding symbol/bit errors. It is therefore of interest to quantify these effects and investigate whether there exist optimal symbol/bit mappings in terms of the SER/BER performance.

\subsection{Design Criteria} \label{sec:method2}

\subsubsection{Minimum SER}

Consider user 1 decoding user 2's message (user 2 decoding user 1's message can be formulated similarly). The average SER at user 1 can be formulated as
\begin{eqnarray}
P_S &=& \sum_{S_1\in {\mathbb Z}_Q} P(S_1)\Bigg[\sum_{S_2\in {\mathbb Z}_Q}P(S_2) \nonumber\\
&& \hspace{-0.4in} \times\sum_{\tilde{S}_2\neq S_2\in {\mathbb Z}_Q} P\Big(M_{R,Q}\big(C(S_1,S_2)\big)\rightarrow M_{R,Q}\big(C(S_1,\tilde{S}_2)\big)\Big)\Bigg] \nonumber\\\label{eq:SER0}
\end{eqnarray}
where $P(S_1)=P(S_2)=1/Q$ is the {\it a priori} probabilities, and $P\big(M_{R,Q}\big(C(S_1,S_2)\big)\rightarrow M_{R,Q}\big(C(S_1,\tilde{S}_2)\big)\big)$ is the pairwise error probability (PEP) of decoding user 2's message as $\tilde{S}_2$ given that user 2's message is $S_2$ and user 1's message is $S_1$. Denote $X=M_{R,Q}\big(C(S_1,S_2)\big)$ and $\tilde{X}=M_{R,Q}\big(C(S_1,\tilde{S}_2)\big)$. It is easy to see that all $X\in {\cal A}_Q^{eq}$ and $\tilde{X}\in {\cal A}_Q^{eq}$ are equiprobable with probability $1/Q$. Using the fact that $M_{R,Q}$ is a one-to-one mapping between $S_2$ and $X$ (and $\tilde{S}_2$ and $\tilde{X}$) given $S_1$, and that any $S_1$ yields the same average PEP due to symmetry, we can reformulate (\ref{eq:SER0}) by changing the indices of summation:
\begin{eqnarray}
P_S &=& \sum_{X\in {\cal A}_Q^{eq}} P(X) \sum_{\tilde{X}\neq X\in {\cal A}_Q^{eq}} P(X\rightarrow \tilde{X}) \nonumber\\
&=& \frac{1}{Q}\sum_{X\in {\cal A}_Q^{eq}} \sum_{\tilde{X}\neq X\in {\cal A}_Q^{eq}} P(X\rightarrow \tilde{X}). \label{eq:SER}
\end{eqnarray}
Note that $P(X\rightarrow \tilde{X})$ can be further expressed by
\begin{equation} \label{eq:SER1}
P(X\rightarrow \tilde{X}) = \sum_{X_R\in {\cal A}_Q^{eq}} P(X\rightarrow X_R)\times P(X_R\rightarrow \tilde{X})
\end{equation}
where $P(X\rightarrow X_R)$ is the probability of deciding on $X_R\in {\cal A}_Q^{eq}$ at the relay given that $X$ should be broadcasted for correct decoding at user 1, which characterizes the effect of noisy transmission in the MA phase on the decoding. Likewise, $P(X_R\rightarrow \tilde{X})$ characterizes the effect of noisy transmission in the BC phase on the decoding.

To measure $P(X\rightarrow X_R)$, it is useful to consider the mapping from the superposed constellation points at the relay $M_Q(S_1)+M_Q(S_2)$, to the denoised signal $C(S_1,S_2)$, and then to the modulated denoised signal $M_{R,Q}\big(C(S_1,S_2)\big)$. We let $M_{R,Q}(U)=W_U\in {\cal A}_Q^{eq}$. As an example, if uniform 4-PAM is adopted in the MA phase, the mapping is given by
\begin{eqnarray}
\{-6,2\}\mapsto 0 \mapsto W_0 \nonumber\\
\{-4,4\}\mapsto 1 \mapsto W_1 \nonumber\\
\{-2,6\}\mapsto 2 \mapsto W_2 \nonumber\\
\{0\}\mapsto 3 \mapsto W_3 \label{eq:mapping_ex}
\end{eqnarray}
where $\{W_0,W_1,W_2,W_3\}$ is a permutation of the elements of ${\cal A}_4^{eq}=\{-3,-1,1,3\}$. Each permutation specifies a symbol mapping at the relay and can be characterized by a $4\times 4$ permutation matrix ${\mathbf P}$ for which $\big[W_0,W_1,W_2,W_3\big]{\mathbf P}=[-3,-1,1,3]$. In general, each symbol mapping $M_{R,Q}$ corresponds to a $Q\times Q$ permutation matrix ${\mathbf P}$ such that $\big[W_0,W_1,\ldots,W_{Q-1}\big]{\mathbf P}=\big[A_0,A_1,\ldots,A_{Q-1}\big]$, where $A_i$ is the $(i+1)$th element of ${\cal A}_Q^{eq}$. Clearly, $P(X\rightarrow X_R), \forall X, X_R\in {\cal A}_Q^{eq}$ are given by $P(W_i\rightarrow W_j), \forall i,j\in {\mathbb Z}_Q$. Obtaining $P(W_i\rightarrow W_j)$ entails the calculation of the probability of the relay detecting a superposed constellation point that maps to $W_j$ given that a superposed constellation point that maps to $W_i$ should be received if no error occurred. For the example in (\ref{eq:mapping_ex}), $P(W_0\rightarrow W_1)$ represents the probability of detecting a superposed constellation point in the set $\{-4,4\}$ given that a superposed constellation point in the set $\{-6,2\}$ should be received if no error occurred. Taking into account the {\it a priori} probabilities of receiving $-6$ and $2$ at the relay, we have
\begin{eqnarray}
P(W_0\rightarrow W_1) &=& P\Big(M_4(\hat{S}_1)+M_4(\hat{S}_2)\in\{-4,4\}\Bigm| \nonumber\\
&& \hspace{0.5in} M_4(S_1)+M_4(S_2)\in\{-6,2\}\Big) \nonumber\\
&=& \frac{1}{4} P\Big(M_4(\hat{S}_1)+M_4(\hat{S}_2)\in\{-4,4\}\Bigm| \nonumber\\
&& \hspace{0.5in} M_4(S_1)+M_4(S_2)=-6\Big) \nonumber\\
&& +\frac{3}{4} P\Big(M_4(\hat{S}_1)+M_4(\hat{S}_2)\in\{-4,4\}\Bigm| \nonumber\\
&& \hspace{0.5in} M_4(S_1)+M_4(S_2)=2\Big) \nonumber\\
&=& \frac{1}{4}\Big(f_{\sigma_1}(-6,-5,-3)+f_{\sigma_1}(-6,3,5)\Big) \nonumber\\
&& +\frac{3}{4}\Big(f_{\sigma_1}(2,-5,-3)+f_{\sigma_1}(2,3,5)\Big)
\end{eqnarray}
where
\begin{equation} \label{eq:f}
f_{\sigma_1}(m,a,b)\triangleq \frac{1}{\sqrt{2\pi\sigma_1^2}}\int_a^b e^{-\frac{(x-m)^2}{2\sigma_1^2}}dx
\end{equation}
and $\sigma_1^2$ is $\sigma^2$ scaled by the average power of the constellation used in the MA phase; in this example, $\sigma_1^2=5\sigma^2$. Following a similar derivation we can obtain all $P(W_i\rightarrow W_j)$. Thus, $P(X\rightarrow X_R)$ for all combinations of $X$ and $X_R$ can be described by a $Q\times Q$ matrix ${\mathbf U}$, where the $(i+1,j+1)$-entry of ${\mathbf U}$ is given by $P(W_i\rightarrow W_j)$. The matrix ${\mathbf U}$ depends on the uniform/nonuniform PAM used in the MA phase.

\begin{figure*}[!bt]
\normalsize
\setcounter{mytempeqncnt}{\value{equation}}
\setcounter{equation}{18}
\begin{eqnarray}
P_B &=& \sum_{S_1\in {\mathbb Z}_Q} P(S_1)\Bigg[\frac{1}{Q\log_2 Q} \sum_{X\in {\cal A}_Q^{eq}} \sum_{\tilde{X}\neq X\in {\cal A}_Q^{eq}} P(X\rightarrow \tilde{X}) \nonumber\\
&& \hspace{1.5in} \times d_H\Big(B_Q(x:M_{R,Q}(C(S_1,x))=X),B_Q(x:M_{R,Q}(C(S_1,x))=\tilde{X})\Big)\Bigg] \nonumber\\
&=& \frac{1}{Q\log_2 Q} \sum_{X\in {\cal A}_Q^{eq}} \sum_{\tilde{X}\neq X\in {\cal A}_Q^{eq}} P(X\rightarrow \tilde{X}) \nonumber\\
&& \hspace{0.5in} \times \Bigg[\underbrace{\sum_{S_1\in {\mathbb Z}_Q} P(S_1)\times d_H\Big(B_Q(x:M_{R,Q}(C(S_1,x))=X),B_Q(x:M_{R,Q}(C(S_1,x))=\tilde{X})\Big)}_{\triangleq N_B(X\rightarrow \tilde{X})}\Bigg] \nonumber\\
&=& \frac{1}{Q\log_2 Q} \sum_{X\in {\cal A}_Q^{eq}} \sum_{\tilde{X}\neq X\in {\cal A}_Q^{eq}} P(X\rightarrow \tilde{X})\times N_B(X\rightarrow \tilde{X}) \label{eq:BER}
\end{eqnarray}
\setcounter{equation}{\value{mytempeqncnt}}
\hrulefill
\vspace*{4pt}
\end{figure*}

Similarly, we can describe $P(X_R\rightarrow \tilde{X})$ for all combinations of $X_R$ and $\tilde{X}$ by a $Q\times Q$ matrix ${\mathbf D}$, where the $(i+1,j+1)$-entry of ${\mathbf D}$ is given by $P(A_i\rightarrow A_j)$. For example, if 4-PAM is used in the BC phase, the $(1,2)$-entry of ${\mathbf D}$ is given by
\begin{equation}
P(-3\rightarrow -1)=f_{\sigma_2}(-3,-2,0)
\end{equation}
where $f_{\sigma_2}$ is defined similarly as $f_{\sigma_1}$ in (\ref{eq:f}) with $\sigma_1$ substituted by $\sigma_2$, where $\sigma_2^2$ is $\sigma^2$ scaled by the average power of the constellation used in the BC phase; in this example, $\sigma_2^2=5\sigma^2$.

It is not difficult to see that $P(X\rightarrow \tilde{X})$ for all combinations of $X$ and $\tilde{X}$ can be described by the matrix product $\big({\mathbf P}^T {\mathbf U} {\mathbf P}\big){\mathbf D}$ according to (\ref{eq:SER1}). As a result, $P_S$ is given by the sum of all off-diagonal entries of ${\mathbf P}^T {\mathbf U} {\mathbf P} {\mathbf D}$ scaled by $1/Q$. Dropping the constant $1/Q$, the optimal symbol mapping $M_{R,Q}$ that minimizes the SER is specified by the solution to the following problem:
\begin{eqnarray}
\min_{{\mathbf P}} && {\mathbf 1}^T \big({\mathbf P}^T {\mathbf U} {\mathbf P} {\mathbf D}\big) {\mathbf 1} - \mbox{Tr}\big({\mathbf P}^T {\mathbf U} {\mathbf P} {\mathbf D}\big) \label{eq:opt_SER}
\end{eqnarray}
where $\mbox{Tr}$ is the trace of a matrix and ${\mathbf 1}$ is a $Q\times 1$ vector in which every element is equal to one. The first term in (\ref{eq:opt_SER}) simply gives the sum of all entries of ${\mathbf P}^T {\mathbf U} {\mathbf P} {\mathbf D}$. The size of the search space is given by the number of possibilities of ${\mathbf P}$, which is $Q!$. The computation can be conducted at the relay.

\subsubsection{Minimum BER}

Similar to the SER formulation in (\ref{eq:SER0}), the average BER at user 1 can be formulated as
\begin{eqnarray}
P_B &=& \sum_{S_1\in {\mathbb Z}_Q} P(S_1)\Bigg[\sum_{S_2\in {\mathbb Z}_Q}P(S_2) \nonumber\\
&& \hspace{-0.5in} \times\sum_{\tilde{S}_2\neq S_2\in {\mathbb Z}_Q} P\Big(M_{R,Q}\big(C(S_1,S_2)\big)\rightarrow M_{R,Q}\big(C(S_1,\tilde{S}_2)\big)\Big) \nonumber\\
&& \hspace{1in} \times\frac{d_H\big(B_Q(S_2),B_Q(\tilde{S}_2)\big)}{\log_2 Q}\Bigg]
\end{eqnarray}
where $d_H\big(B_Q(S_2),B_Q(\tilde{S}_2)\big)$ is the number of bit errors that corresponds to the symbol error $S_2\rightarrow \tilde{S}_2$, and $\log_2 Q$ is the number of bits that represent each source symbol $S_i$. By a similar technique of changing the indices of summation, we can derive (\ref{eq:BER}), shown at the top of this page. In (\ref{eq:BER}), $N_B(X\rightarrow \tilde{X})$ is the average number of bit errors when user 1 detects $\tilde{X}$ given that $X$ should be received if no error occurred. Note that $N_B(X\rightarrow \tilde{X})$ is an average over user 1's message $S_1$. The $N_B(X\rightarrow \tilde{X})$ for all combinations of $X$ and $\tilde{X}$ can be described by a $Q\times Q$ matrix ${\mathbf B}$, where the $(i+1,j+1)$-entry of ${\mathbf B}$ ($i\neq j$) records the average number of bit errors when the error pattern $W_i\rightarrow W_j$ occurs. ${\mathbf B}$ is a symmetric matrix with all diagonal entries equal to zero. Each bit mapping $B_Q$ corresponds to a ${\mathbf B}$. It can be seen from (\ref{eq:BER}) that $P_B$ is given by the sum of all off-diagonal entries of $\big({\mathbf P}^T {\mathbf U} {\mathbf P} {\mathbf D}\big)\circ \big({\mathbf P}^T {\mathbf B} {\mathbf P}\big)$ scaled by $1/(Q\log_2 Q)$, where $\circ$ denotes entrywise product. Dropping the constant $1/(Q\log_2 Q)$ and using the fact that all diagonal entries of ${\mathbf B}$ are equal to zero, the joint optimal symbol mapping $M_{R,Q}$ and bit mapping $B_Q$ that minimizes the BER is specified by the solution to the following problem:
\addtocounter{equation}{1}
\begin{eqnarray}
\min_{{\mathbf P},{\mathbf B}} && {\mathbf 1}^T \Big(\big({\mathbf P}^T {\mathbf U} {\mathbf P} {\mathbf D}\big)\circ \big({\mathbf P}^T {\mathbf B} {\mathbf P}\big)\Big) {\mathbf 1}. \label{eq:opt_BER}
\end{eqnarray}
The size of the search space is given by the number of different joint combinations of ${\mathbf P}$ and ${\mathbf B}$, which is $(Q!)^2$. The computation can be conducted at the relay with the optimal bit mapping results delivered to the two users for employment at the user side.

\subsection{Calculation of ${\mathbf B}$} \label{sec:method3}

To solve (\ref{eq:opt_BER}), we need to obtain ${\mathbf B}$ systematically according to different choices of bit mapping $B_Q$. To calculate the average number of bit errors for error pattern $W_i\rightarrow W_j$ ($i\neq j$) from user 1's perspective, it requires first calculating the number of bit errors due to the symbol error $S_2\rightarrow \tilde{S}_2$ for which $W_i\mapsto (S_1,S_2)$ and $W_j\mapsto (S_1,\tilde{S}_2)$ for some $S_1$, and then averaging the result over all $S_1$. Consider the example of using uniform 4-PAM in the MA phase. By consulting the mapping from the modulated denoised signal $\{W_0,W_1,W_2,W_3\}$, to the denoised signal $C(S_1,S_2)$, and then to the ordered source symbol pair $(S_1,S_2)$, i.e.,
\begin{eqnarray}
W_0\mapsto 0 \mapsto (0,0),(1,3),(2,2),(3,1) \nonumber\\
W_1\mapsto 1 \mapsto (0,1),(1,0),(2,3),(3,2) \nonumber\\
W_2\mapsto 2 \mapsto (0,2),(1,1),(2,0),(3,3) \nonumber\\
W_3\mapsto 3 \mapsto (0,3),(1,2),(2,1),(3,0) \label{eq:mapping_ex2}
\end{eqnarray}
it is clear that the average number of bit errors for $W_0\rightarrow W_1$ (or $W_1\rightarrow W_0$) is given by
\begin{eqnarray}
&& \hspace{-0.3in} \frac{1}{4}\Big[d_H\big(B_4(0),B_4(1)\big)+d_H\big(B_4(3),B_4(0)\big) \nonumber\\
&& \hspace{-0.1in} +d_H\big(B_4(2),B_4(3)\big)+d_H\big(B_4(1),B_4(2)\big)\Big] \label{eq:avg_bit_err_example}
\end{eqnarray}
which is equal to $1$ and $1.5$ for $B_4$ being Gray and binary mappings, respectively. Repeating this procedure for all $W_i\rightarrow W_j$ produces
\begin{equation}
{\mathbf B}=\begin{bmatrix}
0 & 1 & 2 & 1 \\
1 & 0 & 1 & 2 \\
2 & 1 & 0 & 1 \\
1 & 2 & 1 & 0
\end{bmatrix}~~\mbox{   and   }~~
{\mathbf B}=\begin{bmatrix}
0 & 1.5 & 1 & 1.5 \\
1.5 & 0 & 1.5 & 1 \\
1 & 1.5 & 0 & 1.5 \\
1.5 & 1 & 1.5 & 0
\end{bmatrix}
\end{equation}
for Gray and binary mappings, respectively. A low-complexity method for calculating ${\mathbf B}$ is described in Sec.~\ref{sec:method5}.

\subsection{Equivalent Symbol Mappings} \label{sec:method4}

\begin{figure*}[tb]
\begin{center}
\includegraphics[width=1.3\columnwidth]{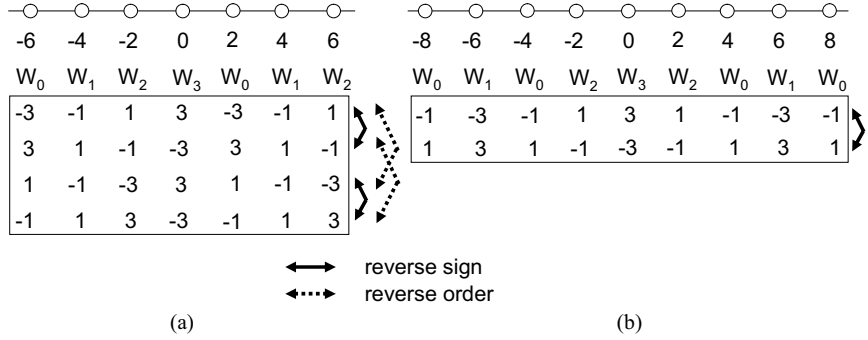}
\caption{Equivalent symbol mappings at the relay for (a) uniform 4-PAM and (b) nonuniform 4-PAM used in the MA phase.} \label{fig:4PAM_symbolmapping}
\end{center}
\end{figure*}

In this and next subsections, we present complexity reduction methods for solving (\ref{eq:opt_SER}) and (\ref{eq:opt_BER}). Essentially, equivalent mappings are identified and the search is restricted to only distinct mappings. Two symbol mappings (specified by different ${\mathbf P}$'s) are equivalent if they yield the same objective value in (\ref{eq:opt_SER}) given ${\mathbf U}$ and ${\mathbf D}$. The equivalent symbol mapping results are numerically verified and conceptually explained as follows for uniform and nonuniform constellations used in the MA phase.

\subsubsection{Uniform PAM}

There are four equivalent mappings for each symbol mapping scheme, as exemplified in Fig.~\ref{fig:4PAM_symbolmapping}(a) for uniform 4-PAM. Mappings with the reverse-order relation yield the same ${\mathbf P}^T{\mathbf U}{\mathbf P}$ and thus the same objective value in (\ref{eq:opt_SER}) due to the symmetric property of the superposed constellation at the relay. Mappings with the reverse-sign relation yield the same sum of off-diagonal entries of ${\mathbf P}^T{\mathbf U}{\mathbf P}{\mathbf D}$ and thus the same objective value in (\ref{eq:opt_SER}) due to the symmetric property of the constellation at the users and at the relay. Due to the equivalent mappings, the size of the search space in (\ref{eq:opt_SER}) is reduced from $Q!$ to $Q!/4$.

\subsubsection{Nonuniform PAM}

There are two equivalent mappings for each symbol mapping scheme, with the reverse-sign relation, as exemplified in Fig.~\ref{fig:4PAM_symbolmapping}(b). The size of the search space in (\ref{eq:opt_SER}) is reduced from $Q!$ to $Q!/2$.

\subsection{Isomorphic and Equivalent Bit Mappings} \label{sec:method5}

Similarly, the size of the search space in (\ref{eq:opt_BER}) can be reduced by restricting the search to only distinct bit mappings. Some bit mappings are isomorphic (e.g., $B_4(\{0,1,2,3\})=\{00,01,11,10\}$ and $B_4(\{0,1,2,3\})=\{00,10,11,01\}$ are both Gray mapping), and some are equivalent in the sense that they yield the same ${\mathbf B}$. It is however not immediately clear how many distinct bit mappings there are for uniform and nonuniform constellations. We propose to first calculate a $1\times Q$ vector ${\mathbf b}$ and then obtain the entries of ${\mathbf B}$ by a table look-up from ${\mathbf b}$. This way, the complexity of calculating ${\mathbf B}$ is reduced, and the number of distinct ${\mathbf B}$'s is simply given by the number of distinct ${\mathbf b}$'s.

\subsubsection{Uniform PAM}

The average number of bit errors for $W_i\rightarrow W_j$ is given by $d_H\big(B_Q(S_2),B_Q(\tilde{S}_2)\big)$ averaged over all $S_1$, where $\tilde{S}_2-S_2=|i-j| ~(\operatorname{mod} Q)$. Thus, we can first calculate ${\mathbf b}=[b_0,b_1,\ldots,b_{Q-1}]$ as
\begin{equation}
b_i=\frac{1}{Q}\sum_{k=0}^{Q-1}d_H\Big(B_Q(k),B_Q\big((k+i)\operatorname{mod} Q\big)\Big), \quad i\in{\mathbb Z}_Q.
\end{equation}
Then, we obtain the $(i+1,j+1)$-entry of ${\mathbf B}$ by a table look-up, i.e.,
\begin{equation}
({\mathbf B})_{i+1,j+1}=b_{|i-j|\operatorname{mod} Q}, \quad i,j\in{\mathbb Z}_Q.
\end{equation}
The complexity of calculating ${\mathbf B}$ this way is in the order of ${\cal O}(2Q^2)$, as opposed to the direct computation in Sec.~\ref{sec:method3} being in the order of ${\cal O}(Q^3)$. Besides, as can be seen, ${\mathbf B}$ will be different if and only if ${\mathbf b}$ is different. The distinct ${\mathbf B}$'s are numerically determined before solving (\ref{eq:opt_BER}). Using uniform 4-PAM as an example, it is known that the four constellation points can be represented by three unique bit mappings up to isomorphism, i.e., $B_4(\{0,1,2,3\})=\{00,01,11,10\}$ (Gray mapping), $B_4(\{0,1,2,3\})=\{00,01,10,11\}$ (binary mapping), and $B_4(\{0,1,2,3\})=\{00,11,01,10\}$ (so-called ``third bit mapping"). These three mappings however produce only two distinct ${\mathbf b}$'s, i.e., ${\mathbf b}=[0,1,2,1]$ for Gray mapping and ${\mathbf b}=[0,1.5,1,1.5]$ for both binary mapping and ``third bit mapping." As a result, there are only two distinct bit mappings out of $4!=24$ possibilities for our consideration, and the size of the search space with respect to ${\mathbf B}$ in (\ref{eq:opt_BER}) can be reduced from 24 to 2. It is numerically confirmed that for uniform 8-PAM there are 46 distinct bit mappings out of $8!=40320$ possibilities.

\begin{table*}[tb]
\begin{center}
\caption{The Relay Symbol Mapper $M_{R,Q}(U)=W_U, U\in{\mathbb Z}_Q$ For Different Modulation Schemes Used in the MA Phase}
\label{tab:simulation} \vspace*{1mm}
\begin{tabular}{|ll||cccccccc|}
\hline
\multicolumn{2}{|l||}{Relay symbol mapping} & $W_0$ & $W_1$ & $W_2$ & $W_3$ &&&& \\\hline
Uniform 4-PAM & 1) & $-3$ & $-1$ & $1$ & $3$ &&&& \\
& 2) & $-3$ & $1$ & $-1$ & $3$ &&&& \\\hline
Nonuniform 4-PAM & 1) & $-1$ & $-3$ & $1$ & $3$ &&&& \\
& 2) & $-3$ & $-1$ & $1$ & $3$ &&&& \\
& 3) & $-3$ & $-1$ & $3$ & $1$ &&&& \\
& 4) & $-3$ & $1$ & $3$ & $-1$ &&&& \\\hline\hline
\multicolumn{2}{|l||}{Relay symbol mapping} & $W_0$ & $W_1$ & $W_2$ & $W_3$ & $W_4$ & $W_5$ & $W_6$ & $W_7$ \\\hline
Uniform 8-PAM & 1) & $-7$ & $-5$ & $-3$ & $-1$ & $1$ & $3$ & $5$ & $7$ \\
& 2) & $-5$ & $1$ & $7$ & $-3$ & $3$ & $-7$ & $-1$ & $5$ \\\hline
Nonuniform 8-PAM & 1) & $-3$ & $-1$ & $-5$ & $-7$ & $3$ & $1$ & $5$ & $7$ \\
& 2) & $-3$ & $-1$ & $-5$ & $-7$ & $3$ & $1$ & $7$ & $5$ \\
& 3) & $-3$ & $-1$ & $-7$ & $-5$ & $3$ & $1$ & $7$ & $5$ \\
& 4) & $-7$ & $-1$ & $-5$ & $-3$ & $5$ & $1$ & $7$ & $3$ \\
& 5) & $-7$ & $-1$ & $-3$ & $-5$ & $5$ & $3$ & $7$ & $1$ \\
& 6) & $-7$ & $-5$ & $7$ & $5$ & $-1$ & $-3$ & $1$ & $3$ \\
& 7) & $-7$ & $-5$ & $-3$ & $-1$ & $7$ & $5$ & $3$ & $1$ \\
& 8) & $-7$ & $-3$ & $-5$ & $-1$ & $5$ & $3$ & $7$ & $1$ \\
& 9) & $-7$ & $-3$ & $-1$ & $-5$ & $5$ & $3$ & $7$ & $1$ \\\hline
\end{tabular}
\end{center}
\end{table*}

\subsubsection{Nonuniform PAM}

Owing to the use of a binary PNC-based denoise mapper, the average number of bit errors for $W_i\rightarrow W_j$ is given by $d_H\big(B_Q(S_2),B_Q(\tilde{S}_2)\big)$ averaged over all $S_1$, where $S_2\oplus_q \tilde{S}_2=i\oplus_q j$. Similarly, we first calculate ${\mathbf b}=[b_0,b_1,\ldots,b_{Q-1}]$ as
\begin{equation}
b_i=\frac{1}{Q}\sum_{k=0}^{Q-1}d_H\Big(B_Q(k),B_Q(k\oplus_q i)\Big), \quad i\in{\mathbb Z}_Q
\end{equation}
and then obtain the $(i+1,j+1)$-entry of ${\mathbf B}$ by a table look-up, i.e.,
\begin{equation}
({\mathbf B})_{i+1,j+1}=b_{i\oplus_q j}, \quad i,j\in{\mathbb Z}_Q.
\end{equation}
For nonuniform 4-PAM, the three unique bit mappings up to isomorphism produce three distinct ${\mathbf b}$'s, namely, ${\mathbf b}=[0,1,2,1]$ for Gray mapping, ${\mathbf b}=[0,1,1,2]$ for binary mapping, and ${\mathbf b}=[0,2,1,1]$ for ``third bit mapping." Thus, there are three distinct bit mappings in this case. For nonuniform 8-PAM, it is numerically shown that there are 175 distinct bit mappings. As can be seen, the number of distinct bit mappings is different for uniform and nonuniform PAM used in the MA phase.

\section{Optimal Mapping Results and Discussions} \label{sec:simulation}

In this section, we present the decoding performance results for different symbol and/or bit mappings. The average symbol power of PAM for all transmissions is normalized to one. The SNR for all transmissions is defined as $1/\sigma^2$. The relay symbol mappings shown in figures are summarized in Table \ref{tab:simulation} and are accordingly referred to in the figures. We examine four transmission scenarios separately: uniform/nonuniform 4/8-PAM for the MA phase, each in combination with uniform PAM of the same cardinality for the BC phase.

\begin{figure}[tb]
\begin{center}
\subfigure[]{
    \label{fig:SER_4PAM_eq}
    \includegraphics[width=\columnwidth]{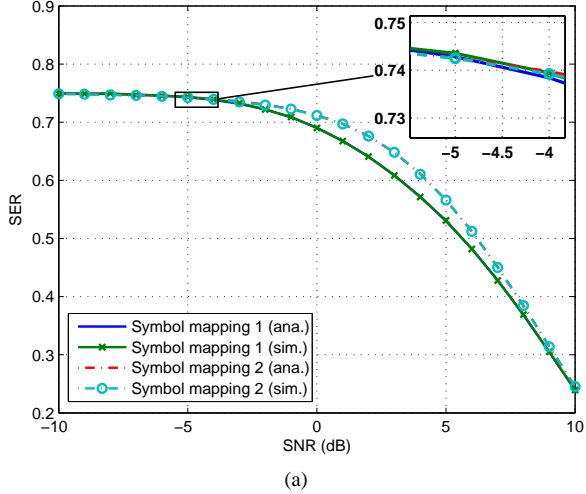}}
\subfigure[]{
    \label{fig:BER_4PAM_eq}
    \includegraphics[width=\columnwidth]{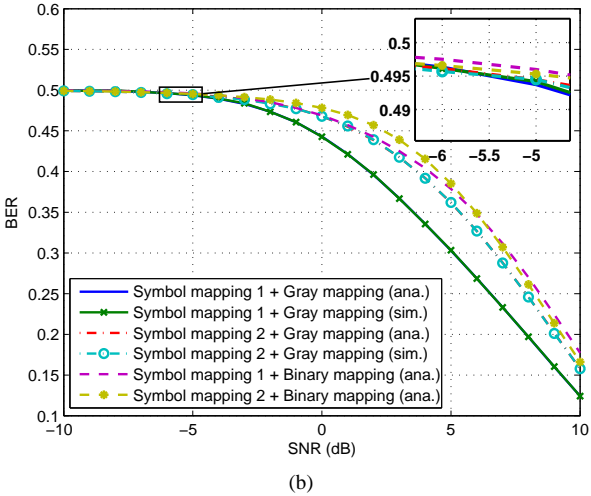}}
\caption{(a) SER performance for different relay symbol mappings and (b) BER performance for different relay symbol mappings in combination with different user bit mappings, for uniform 4-PAM used in the MA phase and uniform 4-PAM used in the BC phase.}
\label{fig:4PAM_eq}
\end{center}
\end{figure}

\subsection{Uniform 4-PAM}

The SER performance for two different symbol mappings is shown in Fig.~\ref{fig:SER_4PAM_eq}. Symbol mapping 2 and symbol mapping 1 are optimal based on the minimum-SER criterion for lower and higher SNR regions, respectively. The difference between the two mappings in the low SNR region is barely discernable and is insignificant from a practical standpoint, with the cross-over numerically verified at around $\mbox{SNR}=[-5,-4]$ dB. Thus, for most practical considerations symbol mapping 1 is the optimal mapping. The SNR dependence of the optimal scheme may be explained by examining the numerical values of ${\mathbf U}$ and ${\mathbf D}$ at different SNRs. For example, for $\mbox{SNR}=10$ dB ${\mathbf U}$ and ${\mathbf D}$ are given by
\begin{eqnarray*}
{\mathbf U} \hspace{-0.1in} &=& \hspace{-0.1in} \begin{bmatrix}
0.862 & 0.079 & 1.9\times 10^{-5} & 0.059 \\
0.079 & 0.843 & 0.079 & 1.1\times 10^{-5} \\
1.9\times 10^{-5} & 0.079 & 0.862 & 0.059 \\
0.079 & 2.2\times 10^{-5} & 0.079 & 0.843
\end{bmatrix},\nonumber\\
{\mathbf D} \hspace{-0.1in} &=& \hspace{-0.1in} \begin{bmatrix}
0.921 & 0.079 & 1.1\times 10^{-5} & 7.7\times 10^{-13} \\
0.079 & 0.843 & 0.079 & 1.1\times 10^{-5} \\
1.1\times 10^{-5} & 0.079 & 0.843 & 0.079 \\
7.7\times 10^{-13} & 1.1\times 10^{-5} & 0.079 & 0.921
\end{bmatrix}
\end{eqnarray*}
and for $\mbox{SNR}=-10$ dB they are given by
\begin{eqnarray*}
{\mathbf U} \hspace{-0.1in} &=& \hspace{-0.1in} \begin{bmatrix}
0.359 & 0.178 & 0.363 & 0.101 \\
0.366 & 0.172 & 0.366 & 0.096 \\
0.363 & 0.178 & 0.359 & 0.101 \\
0.348 & 0.192 & 0.348 & 0.113
\end{bmatrix},\nonumber\\
{\mathbf D} \hspace{-0.1in} &=& \hspace{-0.1in} \begin{bmatrix}
0.556 & 0.108 & 0.096 & 0.240 \\
0.444 & 0.113 & 0.108 & 0.336 \\
0.336 & 0.108 & 0.113 & 0.444 \\
0.240 & 0.096 & 0.108 & 0.556
\end{bmatrix}.
\end{eqnarray*}
As can be seen, the SNR values affect the structures of ${\mathbf U}$ and ${\mathbf D}$ and thus the optimal solution to (\ref{eq:opt_SER}). Specifically, when the channel is more noisy, the ML detected symbol at the user is most likely $-3$ or $3$, and thus ${\mathbf D}$ has large values in the first and fourth columns. When the channel is less noisy, detection errors most likely occur among adjacent symbols, and thus ${\mathbf D}$ has decreasing values away from the correct symbol. Similar SNR dependence is observed in ${\mathbf U}$.

\begin{figure}[tb]
\begin{center}
\includegraphics[width=0.9\columnwidth]{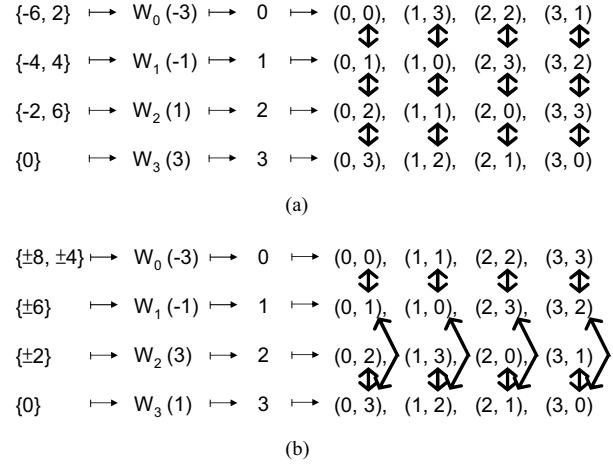}
\caption{The mapping $M_4(S_1)+M_4(S_2)\mapsto\{W_0,W_1,W_2,W_3\}\mapsto C(S_1,S_2)\mapsto (S_1,S_2)$, where the double-ended arrow indicates the dominant symbol error in user 1's decoding of user 2's message ($S_2\leftrightarrow \tilde{S}_2$) given some $S_1$, for (a) uniform 4-PAM and (b) nonuniform 4-PAM used in the MA phase.} \label{fig:4PAM_bitmapping}
\end{center}
\end{figure}

The BER performance for the two symbol mappings in combination with Gray and binary bit mappings is shown in Fig.~\ref{fig:BER_4PAM_eq}. Symbol mapping 2 combined with Gray mapping is optimal for lower SNRs while symbol mapping 1 combined with Gray mapping is optimal for higher SNRs, with the cross-over at around $\mbox{SNR}=[-6,-5]$ dB. Symbol mappings combined with binary mapping (the only distinct bit mapping besides Gray mapping) show various degrees of suffered performance. The results are discussed as follows. As we have seen in Fig.~\ref{fig:errorpatterns}, symbol mapping 2 causes a wider distribution of dominant error patterns than symbol mapping 1, and therefore symbol mapping 1 combined with some judiciously selected bit mapping may yield smaller BER. A simple rule for judicious selection of bit mapping is provided as follows with high-SNR considerations. With symbol mapping 1, the most likely effective errors in the MA phase and the most likely errors in the BC phase are ``aligned" so that they occur only among adjacent constellation points, i.e., $-3$ and $-1$, $-1$ and $1$, and $1$ and $3$. A judicious bit mapping would then be such that $S_2$ and $\tilde{S}_2$ in the dominant decoding errors (from user 1's perspective) $S_2\leftrightarrow\tilde{S}_2=0\leftrightarrow 1, 1\leftrightarrow 2, 2\leftrightarrow 3, 3\leftrightarrow 0$ differ in one bit to minimize the bit errors, as shown in Fig.~\ref{fig:4PAM_bitmapping}(a). This leads to Gray mapping. This analysis suggests an effective symbol and bit mapping design strategy without actually performing the optimization: first choose a symbol mapping that aligns the dominant error patterns in MA and BC phases, and then choose a bit mapping that makes symbols in the dominant error patterns differ in one bit.

\subsection{Nonuniform 4-PAM}

\begin{figure}[tb]
\begin{center}
\subfigure[]{
    \label{fig:SER_4PAM_neq}
    \includegraphics[width=\columnwidth]{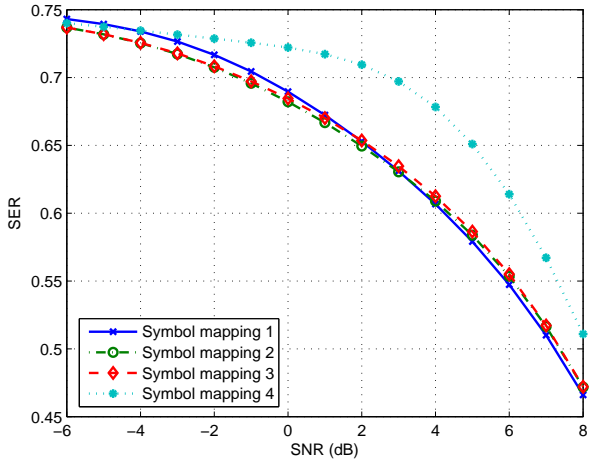}}
\subfigure[]{
    \label{fig:BER_4PAM_neq}
    \includegraphics[width=\columnwidth]{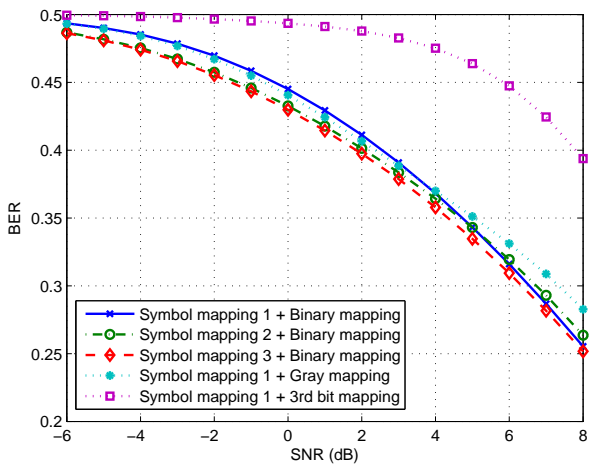}}
\caption{(a) SER performance for different relay symbol mappings and (b) BER performance for different relay symbol mappings in combination with different user bit mappings, for nonuniform 4-PAM used in the MA phase and uniform 4-PAM used in the BC phase.}
\label{fig:4PAM_neq}
\end{center}
\end{figure}

The SER and BER performance results are shown in Fig.~\ref{fig:4PAM_neq}. As seen previously, experimental results closely match the analytical results and thus hereafter only analytical curves are shown in figures for clarity. The optimal symbol mapping based on the minimum-SER criterion is symbol mapping 2 for lower SNRs and symbol mapping 1 for higher SNRs, with the cross-over at around $\mbox{SNR}=[3,4]$ dB. Symbol mapping 3 is suboptimal across all SNRs although the gap to the optimal is small. Symbol mapping 4 exhibits poor performance due to the misalignment between symbol errors in MA and BC phases, and is shown for comparison purposes only. Interestingly, as observed in Fig.~\ref{fig:BER_4PAM_neq}, the combined symbol mapping 3 and binary mapping is optimal based on the minimum-BER criterion across all SNRs even though symbol mapping 3 itself is not the optimal symbol mapping based on the minimum-SER criterion. This shows that the problem in (\ref{eq:opt_BER}) cannot be decoupled into two separate problems including the problem in (\ref{eq:opt_SER}). Symbol mapping 1 in combination with the remaining two distinct bit mapping besides binary mapping (i.e., Gray mapping and ``third bit mapping") show various degrees of degraded BER performance. Comparing Fig.~\ref{fig:4PAM_neq} to Fig.~\ref{fig:4PAM_eq}, nonuniform 4-PAM adopted for the MA phase exhibits worse SER and BER performance in the high SNR region due to the reduced minimum constellation spacing.

The high-SNR analysis presented previously can be applied here to explain the optimal bit mapping result. As shown in Fig.~\ref{fig:4PAM_bitmapping}(b), a judicious bit mapping would be such that $S_2$ and $\tilde{S}_2$ in the dominant decoding errors (from user 1's perspective) $S_2\leftrightarrow\tilde{S}_2=0\leftrightarrow 1, 1\leftrightarrow 3, 2\leftrightarrow 0, 3\leftrightarrow 2$ differ in one bit, which leads to binary mapping.

\subsection{Uniform 8-PAM}

\begin{figure}[tb]
\begin{center}
\subfigure[]{
    \label{fig:SER_8PAM_eq}
    \includegraphics[width=\columnwidth]{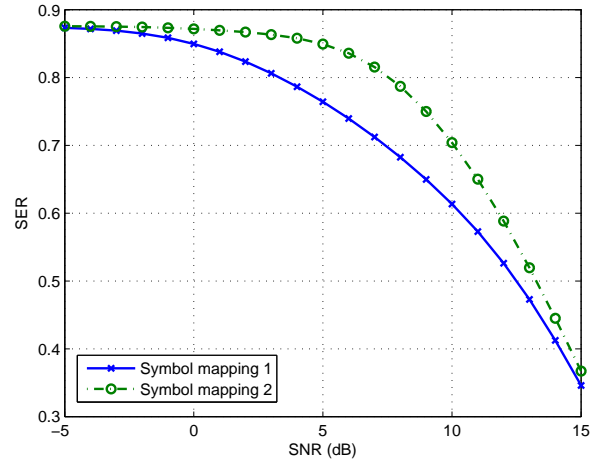}}
\subfigure[]{
    \label{fig:BER_8PAM_eq}
    \includegraphics[width=\columnwidth]{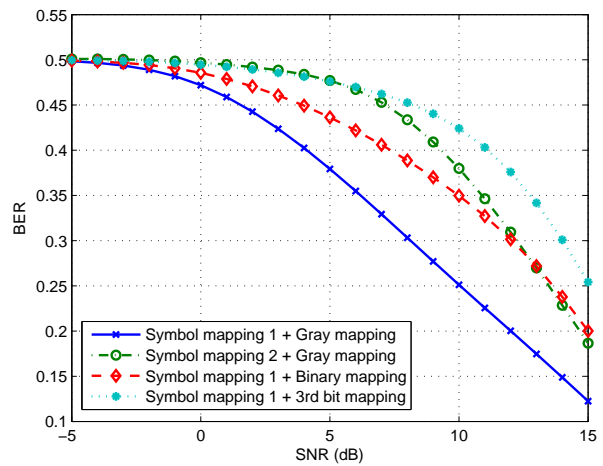}}
\caption{(a) SER performance for different relay symbol mappings and (b) BER performance for different relay symbol mappings in combination with different user bit mappings, for uniform 8-PAM used in the MA phase and uniform 8-PAM used in the BC phase.}
\label{fig:8PAM_eq}
\end{center}
\end{figure}

In Fig.~\ref{fig:SER_8PAM_eq}, we plot the SER performance of the optimal (symbol mapping 1) and the worst-performing (symbol mapping 2) schemes in the higher SNR region based on the minimum-SER criterion. Our numerical results show that the optimal symbol mapping in the lower SNR region varies depending on the SNR although the difference is negligible and practically insignificant. The optimal combined symbol and bit mapping scheme is symbol mapping 1 with Gray mapping $B_8(\{0,1,2,3,4,5,6,7\})=\{000,001,101,100,110,111,011,010\}$ (and its isomorphic variants), as shown in Fig.~\ref{fig:BER_8PAM_eq}. Symbol mapping 1 in combination with two selected distinct bit mappings other than Gray mapping, namely, binary mapping $B_8(\{0,1,2,3,4,5,6,7\})=\{000,001,010,011,100,101,110,111\}$ and ``third bit mapping"\footnote{The ``third bit mapping" here refers to this specific bit mapping adopted in the case of 8-PAM and is not to be confused with the ``third bit mapping" in the case of 4-PAM.} $B_8(\{0,1,2,3,4,5,6,7\})=\{000,011,100,111,010,001,110,101\}$ (and their isomorphic variants), show various degrees of suffered BER performance. Similarities can be drawn with the uniform 4-PAM in terms of the optimal symbol mapping and bit mapping.

\subsection{Nonuniform 8-PAM}

\begin{figure}[tb]
\begin{center}
\subfigure[]{
    \label{fig:SER_8PAM_neq}
    \includegraphics[width=\columnwidth]{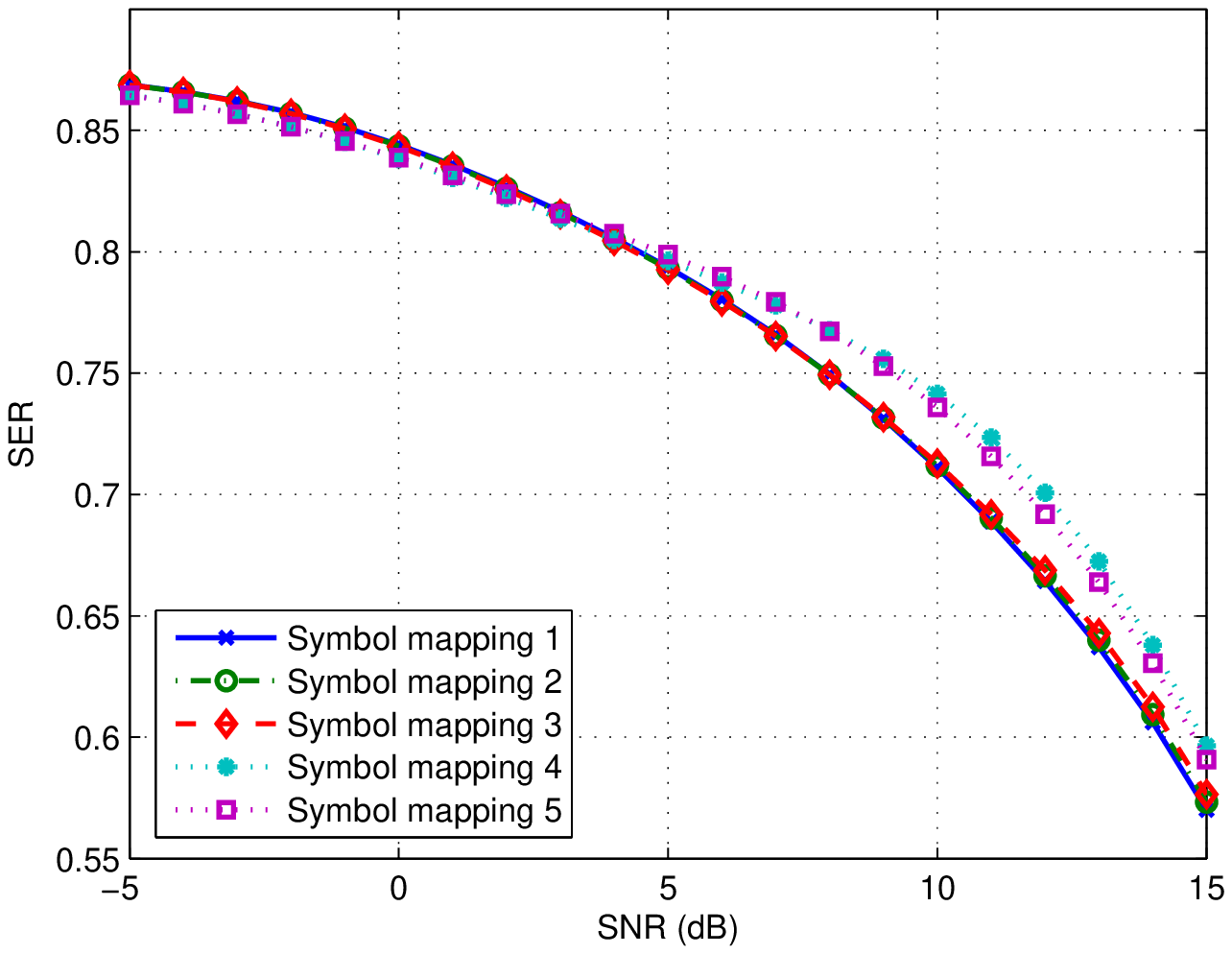}}
\subfigure[]{
    \label{fig:BER_8PAM_neq}
    \includegraphics[width=\columnwidth]{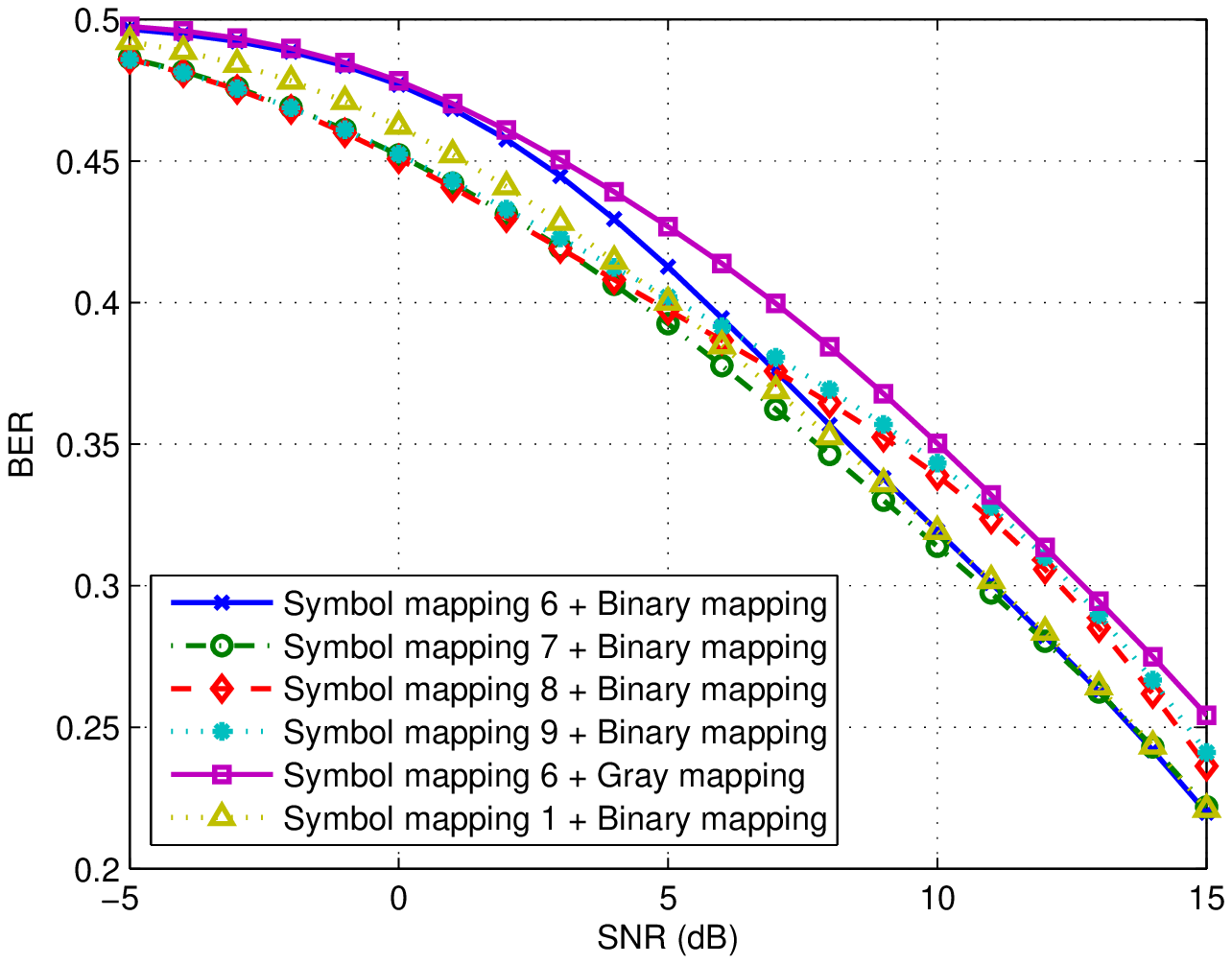}}
\caption{(a) SER performance for different relay symbol mappings and (b) BER performance for different relay symbol mappings in combination with different user bit mappings, for nonuniform 8-PAM used in the MA phase and uniform 8-PAM used in the BC phase.}
\label{fig:8PAM_neq}
\end{center}
\end{figure}

Similar to the case of nonuniform 4-PAM, the SER results in Fig.~\ref{fig:SER_8PAM_neq} demonstrate that the optimal symbol mapping switches among several mapping schemes depending on the SNR. As SNR increases above 4 dB, the optimal symbol mapping switches from symbol mapping 3 to 2 to 1 with negligible differences. As SNR decreases below 4 dB, the optimal symbol mapping switches from symbol mapping 4 to 5 with negligible differences. The more complex mapping result is a consequence of a large number of distinct symbol mappings in this scenario ($8!/2$), some of which produce very similar error probabilities given the 27-level superposed constellation at the relay.

The BER results are shown in Fig.~\ref{fig:BER_8PAM_neq} for several optimal combined symbol and bit mappings depending on the SNR (i.e., symbol mappings 6--9 each combined with binary mapping for $\mbox{SNR}=15; 10 \mbox{ and } 5; 0; \mbox{and } -5$, respectively). As can be seen again, the symbol mapping in the optimal combined symbol and bit mappings based on the minimum-BER criterion is not necessarily the same as the optimal symbol mapping based on the minimum-SER criterion. However, symbol mapping 1, the optimal symbol mapping based on the minimum-SER criterion for higher SNRs, combined with binary mapping exhibits near-optimal BER performance, as shown in Fig.~\ref{fig:BER_8PAM_neq}. This suggests that for practical interest the described high-SNR analysis offers a useful strategy for designing symbol and bit mappings with satisfactory performance, particularly for the case of higher-order PAM. Symbol mapping 6 combined with Gray mapping is also plotted to exemplify the effect of different bit mapping on the system BER performance.

\section{Conclusion} \label{sec:conclusion}

The symbol and bit mapping optimization problem for the DNF communication protocol with PNC for two-way relay networks has been studied. A general design framework was formulated to find the optimal symbol and/or bit mappings. Optimal mapping schemes were presented for both uniform and nonuniform constellations. The main findings are summarized as follows:
\begin{itemize}
\item {\it SNR dependency:} The optimal mappings depend on the SNR of the channel. In particular, the optimal symbol mapping varies across different SNRs for nonuniform 8-PAM due to the large number of distinct symbol mappings.
\item {\it Modulation dependency:} The optimal mappings depend on the modulation used in the MA phase. The different denoise mapper adopted for uniform and nonuniform PAM affects the effective error patterns and thus the optimal mappings. The optimal symbol mapping is one that aligns the error patterns due to noisy transmissions in MA and BC phases, and the optimal bit mapping is Gray mapping and binary mapping for uniform and nonuniform PAM, respectively.
\item {\it Number of distinct symbol and bit mappings:} It is numerically shown that there are $Q!/4$ (or $Q!/2$) distinct symbol mappings for uniform (or nonuniform) $Q$-PAM, and there are 2 (or 3) distinct bit mappings for uniform (or nonuniform) 4-PAM and 46 (or 175) distinct bit mappings for uniform (or nonuniform) 8-PAM.
\end{itemize}

\bibliographystyle{IEEEtran}  
\bibliography{IEEEabrv,mybib}

\begin{IEEEbiography}[{\includegraphics[width=1in,height=1.25in,clip,keepaspectratio]{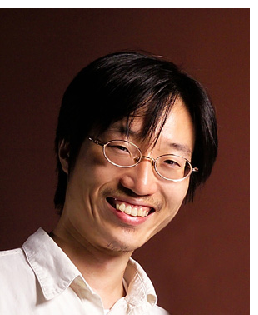}}]
{Ronald Y. Chang} (M'12) received the B.S. degree in electrical engineering from the National Tsing Hua University, Hsinchu, Taiwan, in 2000, the M.S. degree in electronics engineering from the National Chiao Tung University, Hsinchu, in 2002, and the Ph.D. degree in electrical engineering from the University of Southern California (USC), Los Angeles, in 2008. From 2002 to 2003, he was with the Industrial Technology Research Institute, Hsinchu. In 2008, he was a research intern at the Mitsubishi Electric Research Laboratories, Cambridge, MA. Since 2010, he has been with the Research Center for Information Technology Innovation at Academia Sinica, Taipei, Taiwan, where he is now an assistant research fellow. His research interests include wireless communications and networking. He was an Exemplary Reviewer for IEEE Communications Letters in 2012, and a recipient of the Best Paper Award from IEEE Wireless Communications and Networking Conference (WCNC) 2012. He has four awarded and one pending U.S. patents.
\end{IEEEbiography}
\vfill

\newpage

\begin{IEEEbiography}[{\includegraphics[width=1in,height=1.25in,clip,keepaspectratio]{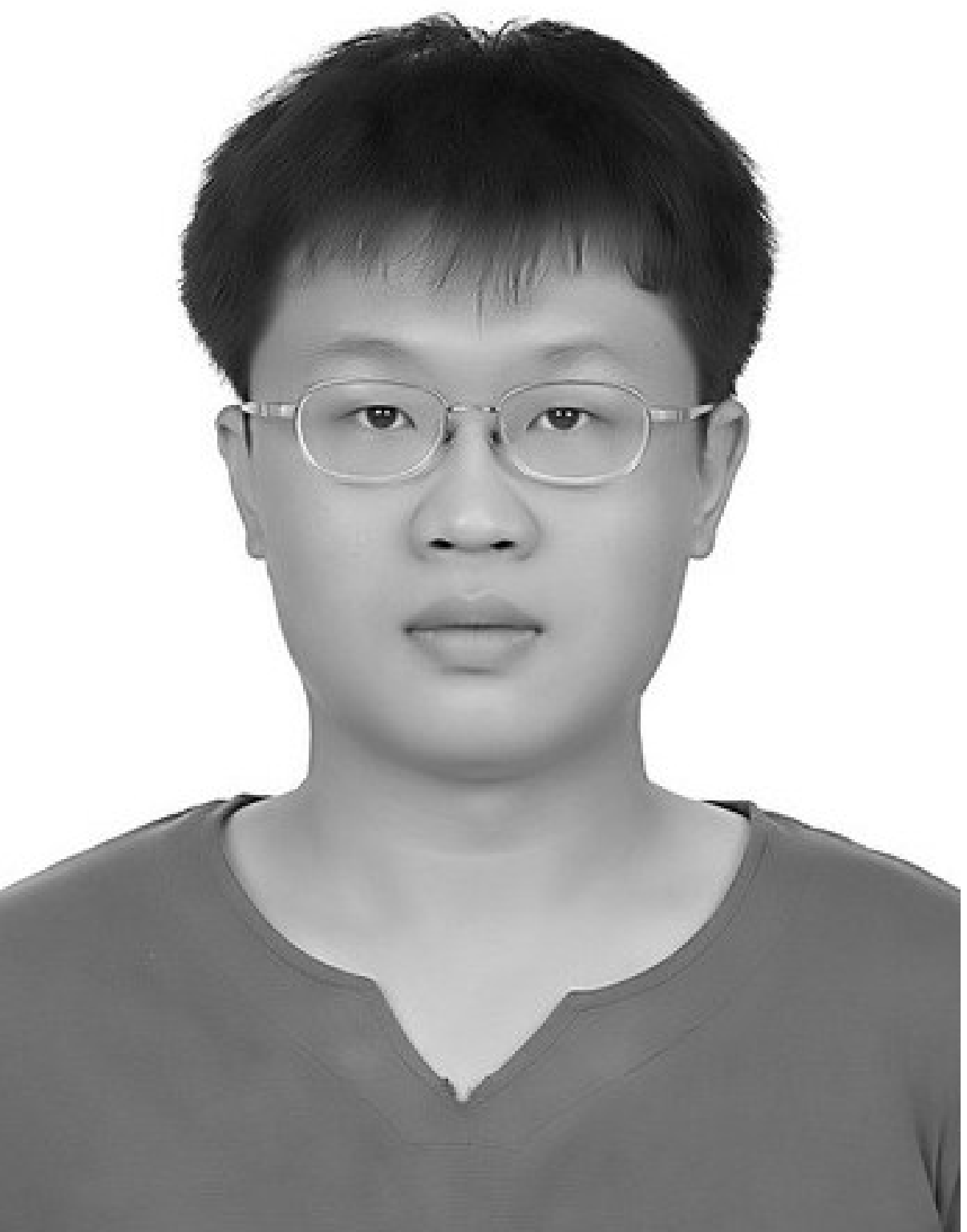}}]
{Sian-Jheng Lin} was born in Taichung, Taiwan, in 1981. He received the B.S., M.S., and Ph.D. degrees in computer science from the National Chiao Tung University, in 2004, 2006, and 2010, respectively. He is currently a postdoctoral fellow with the Research Center for Information Technology Innovation, Academia Sinica. His recent research interests include modulation, data hiding, and error control coding.
\end{IEEEbiography}

\vspace*{-1.5\baselineskip}

\begin{IEEEbiography}[{\includegraphics[width=1in,height=1.25in,clip,keepaspectratio]{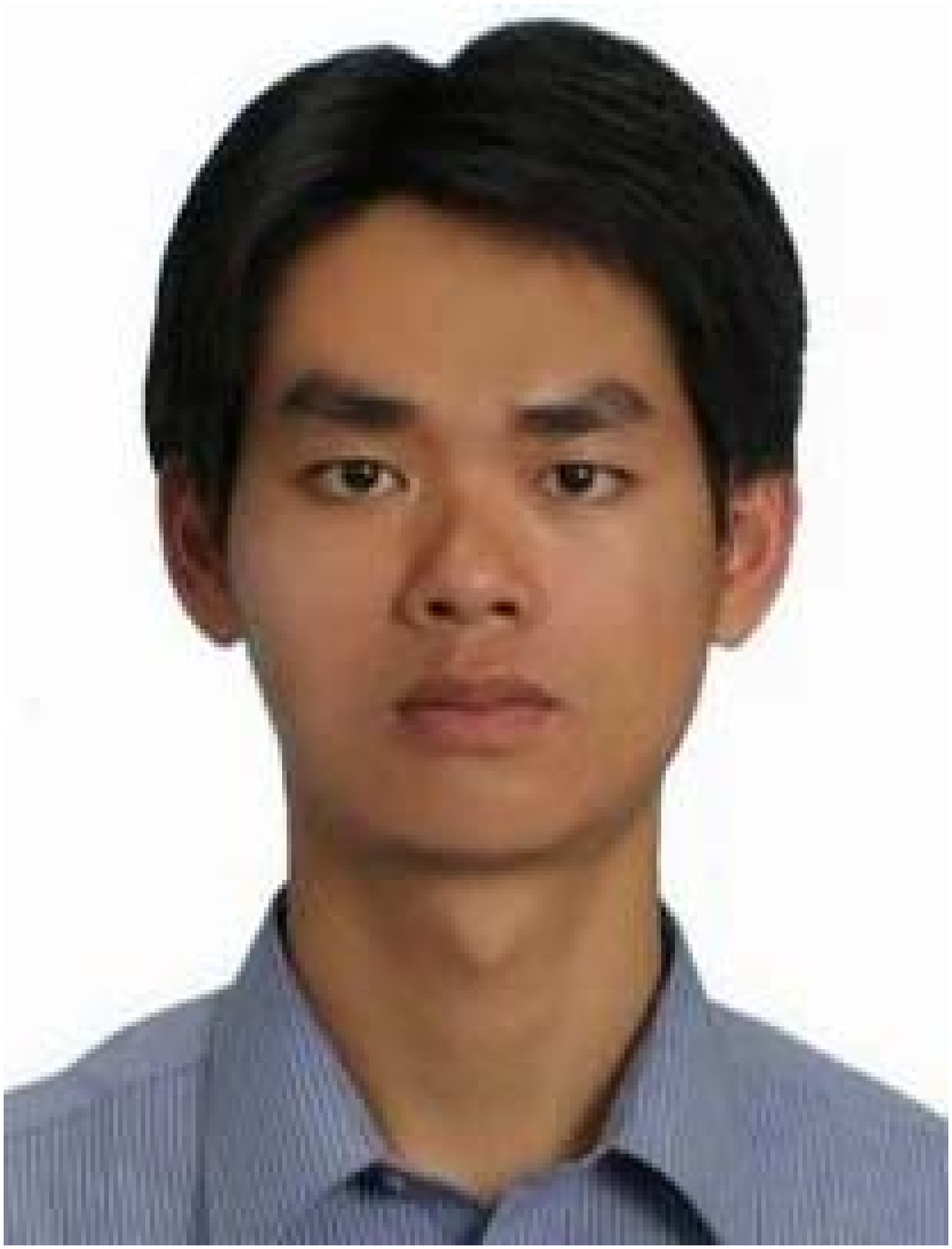}}]
{Wei-Ho Chung} (M'11) received the B.Sc. and M.Sc. degrees in Electrical Engineering from the National Taiwan University, Taipei, Taiwan, in 2000 and 2002, respectively, and the Ph.D. degree in Electrical Engineering from the University of California, Los Angeles, in 2009. From 2002 to 2005, he was a system engineer at ChungHwa Telecommunications Company, where he worked on data networks. In 2008, he was a research intern working on CDMA systems at Qualcomm, Inc., San Diego, CA. His research interests include communications, signal processing, and networks. Dr. Chung received the Taiwan Merit Scholarship from 2005 to 2009 and the Best Paper Award in IEEE WCNC 2012, and has published over 30 refereed journal articles and over 40 refereed conference papers. Since January 2010, Dr. Chung has been a tenure-track assistant research fellow and leads the Wireless Communications Lab in the Research Center for Information Technology Innovation, Academia Sinica, Taiwan.
\end{IEEEbiography}
\vfill

\end{document}